\documentclass{article}
\usepackage{amssymb}
\usepackage{graphicx}

\newcommand{\ba}{\begin{array}}
\newcommand{\ea}{\end{array}}
\newcommand{\bd}{\begin{displaymath}}
\newcommand{\ed}{\end{displaymath}}
\newcommand{\be}{\begin{equation}}
\newcommand{\ee}{\end{equation}}
\newcommand{\bea}{\begin{eqnarray}}
\newcommand{\eea}{\end{eqnarray}}

\begin{document}


\title{{\begin{flushright}
\small{SACLAY-T07/117}\\
\end{flushright}}\vskip 0.5 cm {\textbf {Can Inflation Induce Supersymmetry Breaking in a Metastable Vacuum?}}}
\author{Carlos A. Savoy and Arunansu Sil\\ \\
\small{\em Service de Physique Theorique, CEA/Saclay, F-91191 Gif-sur-Yvette Cedex, FRANCE.}}
\date{}
\maketitle
\begin{abstract}
We argue that fields responsible for inflation  and supersymmetry breaking are 
connected by gravitational couplings. 
In view of the recent progress in studying supersymmetry breaking in a metastable vacuum,
we have shown that in models of supersymmetric hybrid inflation, where 
$R$-symmetry plays an important role, the scale of supersymmetry breaking is 
generated dynamically at the end of inflation and turns out to be consistent 
with gravity mediation.
\end{abstract}
\maketitle
\vskip 0.7cm

Dynamical supersymmetry breaking in a metastable vacuum introduced by Intriligator, 
Seiberg and Shih (ISS)\cite{Intriligator:2006dd} 
is receiving a lot of attention in the recent literature. The ISS model consists in 
\textit{(i)} an asymptotically free SYM theory  with an appropriate number of chiral 
multiplets (metaphorically called ``quarks") which by duality is described by another 
IR free SYM theory at low energy; \textit{(ii)} the addition of a quark mass term in 
the UV (electric) theory. This term is dual to a linear term in the superpotential of 
the IR (or magnetic) theory which has an $R$-symmetry and breaks supersymmetry. The 
non-perturbative dynamical part of this superpotential\cite{ADS} restores 
supersymmetry but leaves a metastable vacuum where it is  broken. There has been 
relevant progress in several directions such as $R$-symmetry breaking\cite{Intriligator:2007py},
\cite{steve}, mediation (basically gauge mediation) of the supersymmetry breaking to the MSSM 
sector\cite{Kitano:2006wm}, gauging of the flavour symmetries\cite{Forste:2006zc} and 
also mechanisms to generate the ISS scale ($\mu_{\rm ISS}$)\cite{Dine:2006gm}.

In this letter, we investigate whether $R$-symmetric gravitational couplings between 
the ISS sector and the sector generating supersymmetric hybrid inflation (SHI)\cite{Dvali:1994ms}, 
would determine the ISS scale. Indeed, we find this scale to be given by the inflation 
scale as fitted in SHI scenarios, $M_{\mathrm{Inf}}$, and the scale of the electric-magnetic 
phase transition of the dual gauge theories, $\Lambda$. Combining an ISS consistency
condition with the fact that $\mu_{\rm ISS}$ is bounded by the EWSB phenomenology  of 
the MSSM we get a relatively narrow allowed interval for $\Lambda$,
hence for $\mu_{\rm ISS}$. It turns out to be consistent with gravity mediation and can be 
made suitable for gauge mediation by a simple change, although the concrete realization 
of the mediation mechanism is not quite addressed here. Notice that the gravitational 
coupling of the two sectors is determined by $R$-symmetry which is instrumental 
in both ISS and SHI mechanisms.

As far as the SHI phenomenology is concerned, without aiming at a careful study of its several 
aspects, the  reheating temperature obtained either by inflaton decay into the quarks of the
ISS sector or through a gravitational coupling between the inflaton and, \textit{e.g.}, right
handed neutrinos, turns out to be phenomenologically adequate. The completion of the model
by the explicit coupling to the MSSM as well as the crucial issue of $R$-symmetry breaking are 
postponed to future publications.

The set-up consists of three components namely, the inflationary sector (Inf), the supersymmetry 
breaking sector (here the ISS sector) and the MSSM sector. The Inf sector consists of superfields 
used to implement inflation. 
The scenario is organized within the framework of the well-known supersymmetric hybrid 
inflation model\cite{Dvali:1994ms}, with a superpotential given by\footnote{Since eq.(\ref{eqn1}) 
has to be quadratic in $\chi$, one can invoke a $Z_2$: $\chi \rightarrow -\chi$, present in 
eq.(\ref{eqn1}) or, as more frequently done in the literature, to introduce a $U(1)$ and a conjugate 
pair $\chi, \bar \chi$. In our case, the $U(1)$ breaking by the $\chi$ vev would yield a goldstone 
boson that would restore supersymmetry in ISS. Our choice here is the simplest one, an alternative 
being to gauge the $U(1)$.}
\be
W_{\rm Inf} = kS(\chi^2 - M_{\rm Inf}^2),  
\label{eqn1}
\ee
where $S, \chi$ are chiral superfields. A $U(1)_R$ symmetry is present under which $S$ has $R$-charge 2 
as well as the superpotential and $\chi$ has no $R$-charge. 
The parameters $k$ and $M_{\rm Inf}$ can be made real and positive by field 
redefinitions\footnote{Alternatively this sector can also be interpreted as strongly coupled supersymmetric 
gauge theories with quantum moduli spaces\cite{Dimopoulos:1997fv}.}.
The interest of the SHI is inherent in its $R$-invariance. 
It has the advantage of avoiding the large supergravity corrections with canonical Kahler potential 
due to the linearity in the $S$ superfield\cite{Dvali:1994ms}.
    
The ISS sector is described by a supersymmetric $SU(N_c)$ gauge symmetry with $N_f$ flavors of quark, antiquark 
pairs in the electric theory. Here $\Lambda$ is the strong-coupling scale of the theory, 
below which the theory can be described as the magnetic dual, $SU(N)$ gauge theory, where 
$N = N_f - N_c$ with $N_f$ flavors of magnetic quarks, $q^c_i, \tilde q^i_c$, ($i = 1, ..,N_f$ and 
$c = 1, ..,N$) and a $N_f \times N_f$ gauge singlet superfield $\Phi^i_j$ (the 
meson field $\Phi = Q \tilde Q / \Lambda$). The magnetic theory is  
IR free if $N_c + 1 \le N_f \le \frac{3}{2}N_c$ and has the superpotential given by 
\be
W_{\rm ISS} = h{\rm Tr}q \Phi \tilde q - h\mu^2_{\rm ISS} {\rm Tr} \Phi,
\label{eqn2}
\ee 
along with the dynamical superpotential 
\be
W_{dyn} = N\left(h^{N_f} \frac{det \Phi}{\Lambda^{N_f - 3N}}\right)^{\frac{1}{N}},
\label{eqn3}
\ee
where $h = O(1)$ and $\mu_{\rm ISS} \ll \Lambda$ are constants. Eq.(\ref{eqn2}) has an 
$R$-symmetry with $R_{\Phi} = 2$ ($R_Q = R_{\tilde Q} = 1$ upto a baryon number) and 
$R_{q, \tilde q} = 0.$ Our aim is to generate the scale of supersymmetry breaking 
$\mu_{\rm ISS}$ (from inflation).
Note that by duality the second term in eq.(\ref{eqn2}) corresponds to a mass term 
$\mu^2_{\rm ISS} Q \tilde Q / \Lambda$ in the electric theory. 
 
We have assumed that the two sectors Inf and ISS can communicate with each other 
only through gravity. This gravitational couplings must have $R$-charge
2 as both $W_{\rm ISS}$ and $W_{\rm Inf}$ have. Therefore all these couplings have 
to be linear in $Q \tilde Q$. The lowest dimensional operator is then given by
\be
W_{int} = \frac{g}{M_P} \chi^2 {\rm Tr}{Q \tilde Q},
\label{eqn4}
\ee 
where $M_P = 2.4 \times 10^{18}$ GeV is the reduced Planck scale and $g$ is a coupling constant.    
Once $\chi$ acquires a vev after inflation, this term will automatically generate the electric 
quark's mass hence its dual, the $\mu_{\rm ISS}$ term of the ISS.       

Now consider the scalar potential obtained from eqs.(\ref{eqn1}), 
(\ref{eqn2}), 
\be
V = k^2|\chi^2 - M^2_{\rm Inf}|^2 + 4|\chi|^2(k^2|S|^2 + g^{2}|Q \tilde Q|^2/M^2_{P}).
\label{eqn5}  
\ee
To realize inflation, $S$ is displaced from its present day location to values that 
exceed $M_{\rm Inf}$. The field $\chi$ is then attracted to the origin by a large mass 
term and the potential is completely flat along $S$. The appearance of a vacuum energy 
density of order $k^2 M^4_{\rm Inf}$ is responsible for inflation. 
The supersymmetry breaking by this vacuum energy can be exploited 
to generate a slope along the inflationary valley ($\chi = 0, |S| > S_c $). 
The existence of the mass splitting in $\chi$ supermultiplet leads to the one 
loop correction\cite{Dvali:1994ms} 
\be
\Delta V = \frac{k^4 M^4_{\rm Inf}}{8 \pi^2} \left[{\rm {ln}}{\frac{2k^2 S^2}{\mu^2_r}} + 
O(M^4_{\rm Inf}/S^4)\right],
\ee
which can be calculated from the Coleman-Weinberg formula. Here $\mu_r$ is a renormalization scale. 
One can then calculate the $\epsilon$ (=$\frac{M^2_P}{2}(\frac{V'}{V})^2$) 
and $\eta$ (= $M^2_P \frac{V''}{V}$) parameters of inflation. The inflation ends when 
$S_c \simeq M_{\rm Inf}$ so that the mass term of $\chi$ becomes negative. The spectral index 
in this class of models is estimated to be $n_s \gtrsim 0.98$, which is only consistent with the 
current WMAP measurement of spectral index 
in the 2$\sigma$ range\cite{Spergel:2006hy}, \cite{Kinney:2006qm}. The scenario 
can be well improved by adding a non-canonical Kahler term\footnote{Another way for lowering the spectral index is discussed in
\cite{lyth}.} as $k_{ss} |S|^4 / M^2_P$ in the 
Kahler potential\cite{BasteroGil:2006cm}. It is shown in \cite{ur Rehman:2006hu} (see Fig.3 therein), that 
with $k_{ss} = 0.01$, the spectral index fit in with the preferred 1$\sigma$ range 
from recent WMAP data, $n_s \simeq 0.95^{+0.015}_{-0.019}$,  where the coupling 
$3 \times 10^{-3} \lesssim k \lesssim 6 \times 10^{-2}$.
At the end of inflation, the Inf system rolls towards the minimum at $V_1$, $S = 0, 
\vert\langle \chi \rangle \vert = M_{\rm Inf}$. 
Now we can realize the impact of the term $W_{int}$. Once the 
$\chi$ field starts acquiring a vev, this term would generate a dynamical mass term for quarks, 
$m_{Q} = g {\langle \chi \rangle}^2 / M_P \ll \Lambda$.          
So at this point the ISS sector can be described by the IR magnetic phase
\be
W_{\rm ISS} = \Phi_{ij} q_i \tilde q_j - m_{Q}\Lambda Tr\Phi + W_{dyn}, 
\ee
which consists of metastable supersymmetry breaking vacua at $ \langle q \rangle = 
\langle \tilde q^T \rangle = \mu_{\rm ISS} = 
\sqrt{m_{Q} \Lambda}, \Phi = 0$.

Therefore we find that the evolution of the ISS system is eventually connected with the dynamics of 
the $\chi$ field. During inflation, $Q, \tilde Q$ acquire positive mass square terms $O(H^2)$ 
($H^2 = k^2 M^4_{\rm Inf}/{3M^2_{P}}$) from supergravity corrections and thereby they settle 
at the origin\footnote{Here we are approximating the  UV Kahler  potential for $Q, \tilde Q$,
with characteristic scale $M_{P}$,
by its canonical form since the relevant UV scale is $M_{\rm Inf} \ll M_{P}$.}. Since the 
relevant inflation scale and $\Lambda$ are not very far away, it is
equally interesting to consider the scheme in the magnetic phase.
Analogously,  one expects  $\Phi$ to get a mass  $O(H)$ during inflation, under the assumption of 
regular enough Kahler potential and to become small enough\footnote{In this 
phase, the metric has an unknown dependence on the scale $\Lambda$ and we assume
the moduli space to be smooth enough around origin. Then the Kahler potential 
is regular there and is given by $K = \alpha Tr \Phi^{\dagger} \Phi + \beta Tr q^{\dagger} q
 + ... $, with $\alpha$ and $\beta$ positive, for $\Phi \ll \Lambda$.  
Another positive contribution to $V (\Phi)$  come  from $W_{dyn}$ but, for $\Phi < \Lambda$, they
at most comparable with the contributions from  $\Phi$ dependent terms in the Kahler potential.}. 
So a nice feature of the model is that either way, we end up at the origin. 
Note that at this time there is no other mass term for $Q, \tilde Q$ (or linear term in $\Phi$)
from $W_{int}$ as $\chi = 0.$     
Once the inflation ends, the Inf system falls toward the minimum at $V_1$ and performs damped 
oscillation about it. On the other hand, when $\chi$ starts to become non-zero after inflation
the term $\mu_{\rm ISS}$ is generated. Taking into account the non-perturbative term, $W_{dyn}$,
it develops the supersymmetric minimum in the ISS sector
at 
\be
\langle q \rangle = \langle \tilde q^T \rangle = 0; ~~ \langle \Phi \rangle = 
\mu_{\rm ISS}(\chi) \left( \epsilon^{\frac{N_f - 3N}{N_c}}\right)^{-1} {\mathbb I}_{N_f},
\ee  
where 
\be
\epsilon = \frac{\mu_{\rm ISS}(\chi)}{\Lambda} \ll 1.
\ee
The other minimum for ISS is at $\Phi = 0$ and $q = \tilde q = \mu_{\rm ISS}$ and becomes a local minimum 
which breaks supersymmetry due to the rank conditions. Here we see that $\Phi$ is situated 
at the origin from the beginning and now there is a possibility that it could end up in the 
supersymmetric minimum. But this is not the case in this scenario. The authors in \cite{Intriligator:2006dd}
have estimated the 
tunneling rate from the supersymmetry breaking to the supersymmetry preserving vacuum 
and the action of the bounce solution is of the form 
\be
S_{bounce} = \frac{2 \pi^2}{3} \frac{N^3}{N^2_f}\left( \frac{\langle \Phi \rangle}{\mu_{\mathrm{ISS}}}\right)^4 
\simeq \frac{1}
{\epsilon^{4(N_f - 3N)/(N_f - N)}} \gg 1, ~~{\rm {for}} ~~\epsilon \ll 1, 
\ee    
since $\chi \lesssim M_{\mathrm{ ISS}}$.
Therefore once the $\Phi$ field is pushed to the origin during 
inflation, it will stay there and that becomes the metastable supersymmetry breaking minimum. In 
other words, our scenario provides a natural explanation why \cite{extra14} the ISS system should be in the 
metastable minimum, not in the supersymmetric minimum.

An inflationary scenario would be complete by a successful reheating process\cite{Kofman:1994rk}.
The superpotential $W_{\rm Inf}$ leads to the common inflaton-system mass as
$m_S = m_{\chi} = 2 k M_{\rm Inf}$. So that when $\chi$ is performing oscillations 
around the minima $V_1$, $\chi$ could decay into ISS quarks/squarks with
\be
\Gamma \simeq \frac{g^2k}{4\pi} \frac{M^3_{\rm Inf}}{M^2_P}, 
\ee
and reheat temperature $O(10^{9-10} ~\rm GeV)$ which is consistent with the gravitino problem\cite{Fujii:2003nr}. 
This is not of great interest from the point of 
view of creating MSSM particles after inflation. Thus we can think off adding  some other 
couplings in $W_{\rm Inf}$, e.g.: (i) $Sh_1h_2$ where $h_{1,2}$ are MSSM higgses carrying zero 
$R$-charge\cite{Dvali:1997uq}; (ii) $\chi^2 h_1 h_2 / M_P$, where $h_{1,2}$ have $R=1$ each; 
or (iii) $f_{ij}\chi^2 N_i N_j / M_P$, where $N_i$ are neutrino superfields and $i, j$
are generation indices. But, (i) will not work in this scenario as it restores 
supersymmetry\footnote{Since R-symmetry cannot forbid a term like 
$Tr{Q \tilde Q}h_1 h_2 / M_P$ thereby spoiling the existence 
of metastable vacua in ISS sector.} while (ii) is also not good as it yields a large 
$\mu$-term.  Instead, (iii) works fine with a reheat temperature $O(10^{9-10} ~\rm GeV)$. 
First of all it provides mass for the right handed neutrinos
$f_{ij} M^2_{\rm Inf}/M_P = O(10^{10} ~\rm GeV)$ (at least one has to be less than $m_{\chi}/2$ for 
$\chi$ to decay) which is not only in the right ballpark to explain the light 
neutrino mass through the see-saw mechanism but also opens up the possibility 
to have non-thermal leptogenesis\cite{Fukugita:1986hr}.

Now we want to evaluate the possible constraints over the mass scales  
$\mu_{\rm ISS}, \Lambda$.  They are:\\ 
\noindent (a) metastability condition: ~~$\mu_{\rm ISS} \ll \Lambda$ ~to preserve the ISS vacuum,\\
\noindent (b) supersymmetry mediation condition: ~~$m_{\mathrm{susy}} M_P \simeq F_{sugra} \ge F_{ISS} = \mu^2_{\rm ISS}$, \\
\noindent where $m_{\mathrm{susy}} = O(\rm TeV)$ is the order of magnitude of the soft masses in the effective MSSM 
lagrangian and (b) means that the mediation scale has to be less than the Planck mass.
The condition (a) and (b) respectively translate into:
\be
 \Lambda \gg \frac{g}{k} \frac{\sqrt{V_{\rm Inf}}}{M_P},~~ {\rm and} ~~ \Lambda \le \frac{k}{g} 
\frac{M^2_P}{\sqrt{V_{\rm Inf}}}m_{\mathrm{susy}}.
\label{condi}
\ee
From the recent analysis done in \cite{ur Rehman:2006hu} (See Fig.4 therein),
we get ${V^{1/4}_{\rm {Inf}}} = \sqrt{k} M_{\rm Inf}$ lies between $2.0 \times 10^{13}$ and $10^{15}$ 
GeV which corresponds to the spectral index $n_S$ in the 1$\sigma$ range of WMAP3 data 
(with the same $k_{ss} = 0.01$ as we have considered earlier). 
Therefore the conditions in eq.(\ref{condi}) can be simultaneously satisfied in the lower 
range for $V_{\rm {Inf}} = O({10^{13 - 14}} ~{\rm GeV})^4$. 
Without detailed study of the parameter space of the two independent couplings $k$ and $g$,
we see that with $k = O(10^{-2})$ and $g = O(10^{-1} - 10^{-2})$ these conditions meet leading to 
$\mu_{\rm ISS} = O(10^{12} ~\rm GeV)$ and $\Lambda = O(10^{14} ~\rm GeV)$. 

With this order of magnitude for the 
supersymmetry breaking scale, supergravity mediation could be sufficient 
to give mass to scalars. But this presupposes a cosmological constant suppression mechanism
is at work so to cancel  the $V = O(\mu_{\rm ISS}^{4})$ contribution from (\ref{eqn2})
and (\ref{eqn3}) - which in any instance has to be made consistent with data on the 
cosmological constant\footnote{It is well-known that gravity mediation assumes this cancellation 
which relates the auxiliary field that break supersymmetry and the gravitino mass. Instead, here, 
the superpotential vanishes at $\Phi=0$  so that the model should be further 
elaborated.}. However, it can be adapted to yield lower values of  $\mu_{\rm ISS}$,
possibly consistent with, {\it e.g.}, gaugino mediation, by modifying the dependence of 
(\ref{eqn1}) on $\chi$ so  that $\mu_{\rm ISS}$ would be reduced by powers of 
$M_{\rm Inf}/M_P$; for instance, a term $\chi^{4}$ which could also improve the fit to the 
cosmological parameters.  Of course a supersymmetry breaking mediation mechanism should
be concocted\footnote{Work in progress in collaboration with Philippe Brax.}. 

Gaugino mass generation remains a problem since it is closely related to $R$-symmetry breaking 
while $\Phi = 0$ at the metastable vacuum. The non-perturbative term, $W_{int}$ explicitly 
breaks $R$-symmetry and produces the supersymmetry preserving (and $R$-symmetry 
breaking) vacuum. The Coleman-Weinberg correction may shift $\Phi$ from origin 
which breaks $R$-symmetry\footnote{Once $R$-symmetry is broken spontaneously, it leads to the 
R-axion problem. As we are in the framework of supergravity, it is possible to make R-axions 
sufficiently heavy\cite{bagger} due the explicit breaking of R-symmetry 
via the constant term present in the superpotential in order to get a realistic cosmological constant.}, 
still all the fields have $R=0,2$, and gaugino can not get mass\cite{Intriligator:2007py},
the direct coupling of $\Phi$ to the gauginos is forbidden by $R$-symmetry. An approach 
towards solving this $R$-symmetry problem is recently discussed recently in \cite{steve}. 
In this letter we do not address the possibility of embedding our scenario into that kind of model to 
get a better phenomenologically viable model and we keep it for future work. 

In conclusion we have given a simple model to naturally provide the scale
of supersymmetry breaking from inflation in the context of ISS model. Quarks become massive after
inflation and the system ends up in the metastable supersymmetry breaking vacuum. We also
realize that the scale of supersymmetry breaking (which is related with the scale of inflation)
falls in a range where gravity mediation of the supersymmetry breaking into MSSM sector is possible.

\vskip 10pt
    
\noindent {\bf \textit{Acknowledgements}} -  This work is supported by 
the RTN European Program MRTN-CT-2004-503369
and by the French ANR Program PHYS@COL\&COS.
The authors thank Steven Abel and Philippe Brax for stimulating discussions
and the referee for interesting comments.

\end{document}